\documentclass[a4paper]{jpconf}
\usepackage[utf8]{inputenc}
\usepackage{graphicx}
\usepackage{amsmath}
\usepackage{listings}
\usepackage{xcolor}

\usepackage{subcaption}
\begin{document}
\title{An automated framework for hierarchical reconstruction of $B$ mesons at the Belle~II experiment}

\author{C Pulvermacher, T Keck, M Feindt, M Heck and T Kuhr}
\address{
Institut für Experimentelle Kernphysik,
Karlsruhe Institute of Technology (KIT),
Wolfgang-Gaede-Str. 1,
76131 Karlsruhe, DE}

\ead{christian.pulvermacher@kit.edu}

\begin{abstract}

We present a software framework for Belle II that reconstructs $B$ mesons in many decay modes with minimal user intervention.
It does so by reconstructing particles in user-supplied decay channels, and then in turn using these reconstructed particles in higher-level decays.
This hierarchical reconstruction allows one to cover a relatively high fraction of all $B$ decays by specifying a limited number of particle decays.
Multivariate classification methods are used to achieve a high signal-to-background ratio in each individual channel.
The entire reconstruction, including the application of pre-cuts and classifier trainings, is automated to a high degree and will allow users
to retrain to account for analysis-specific signal-side selections.

\end{abstract}

\section{Introduction}
Belle II is an experiment being built at the $e^+e^-$ SuperKEKB $B$ factory in Tsukuba, Japan, 
and is planned to have a luminosity 40 times higher than its predecessor Belle~\cite{tdr}.
Like other $B$ factories the collider will operate mainly on the energy of the $\Upsilon(4S)$ resonance, which practically always decays into pairs of $B$ mesons.
Since these events consist of only $B^+ B^-$ or $B^0 \bar B^0$ pairs and their decay products---with only minor contamination from beam or electronics background---they enable the use of reconstruction techniques that would not be available at a hadron collider.

Specifically, reconstructing one $B$ meson (called $B_{\text{tag}}$) allows one to infer information about the other $B$, including its four-momentum, without explicitly reconstructing it.
Combined with a signal selection, the additional information about $B_\text{sig}$ allows to improve the reconstruction, e.g. in channels with neutrinos.
Additionally, one can also make use of the fact that in a correctly reconstructed event every track must come from one of the $B$ mesons, so requiring the absence of additional particles improves the purity of the selection.
This is equivalent to reconstructing the $\Upsilon(4S)$, but of course requires a high efficiency in the $B_\text{tag}$ reconstruction to be of use.

\section{Hierarchical reconstruction}
\begin{figure}
\begin{center}
\includegraphics[width=0.9\textwidth]{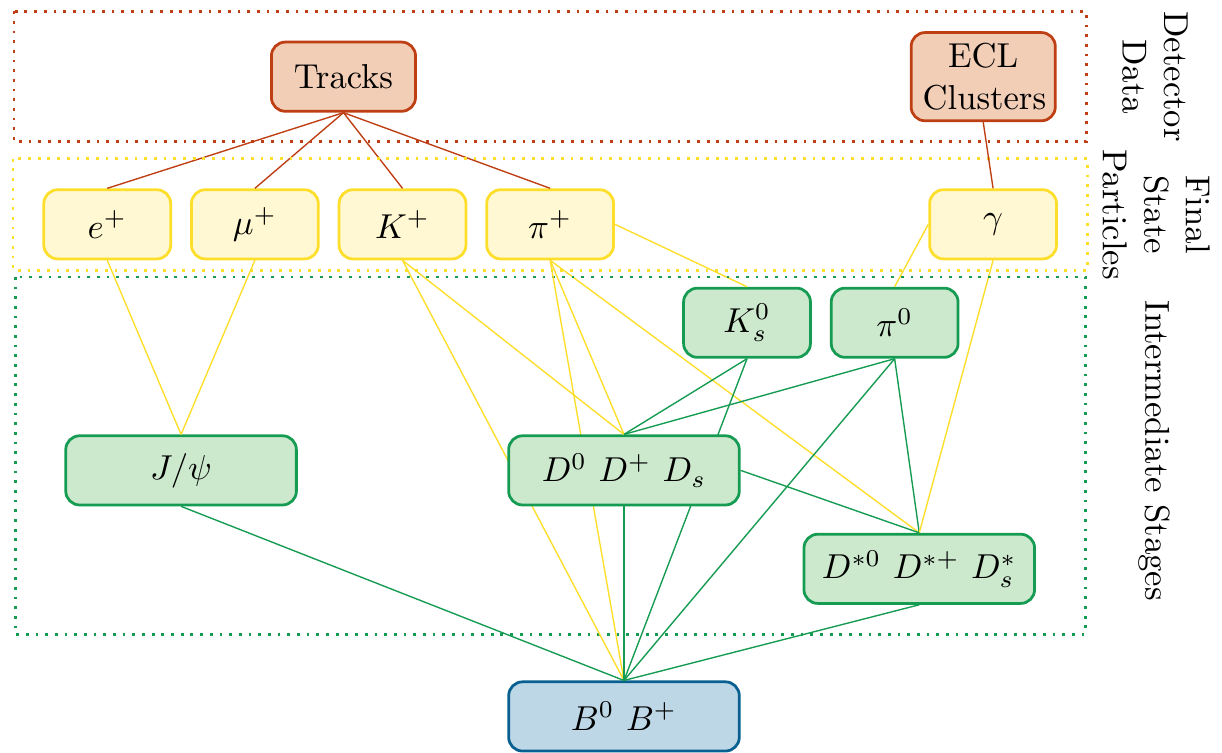}

\end{center}
\caption{\label{network}Illustration of the hierarchical reconstruction from final state particles to $B$ mesons.}
\end{figure}

To this end, a neural-network-assisted reconstruction of $B$ mesons in many decay channels was pioneered at the Belle experiment which reconstructs decays of particles into their immediate daughters and combines the output until reaching the $B$ meson level~\cite{fullreco}.
This principle is illustrated in Figure~\ref{network}.
To reduce the computational burden imposed by the combinatorics of the reconstruction, soft cuts based on the network output remove background at each stage without overly compromising efficiency.
The network output can be interpreted as a probability for the candidate to be correctly reconstructed, which improves its usefulness as an input for subsequent networks even though candidates can originate from a number of different decay channels.

Compared to an explicit reconstruction of each exclusive decay channel, this hierarchical approach greatly reduces the number of multivariate classifiers needed and consequently increases the statistics in each of these channels.
For the Full Reconstruction method used at Belle, about 1000 exclusive decay channels are handled by $\mathcal{O}(100)$ artificial neural networks.
The efficiency of this method is about 0.18\,\% for $B^0$ mesons and 0.28\,\% for $B^\pm$ mesons, i.\,e. for the final Belle $\Upsilon(4S)$ data sample containing $771.6 \times 10^6$ $B\bar B$ pairs 3.5 million tag-side $B$ mesons are correctly reconstructed.

For Belle~II, we want to further increase the efficiency, by being able to easily add additional decay channels and by optimising cuts and multivariate methods.
To gain the most benefit from this, the process should be automatic to a significant degree,
so that after a user adds or modifies a decay channel, the training of classifiers and determination of applied cuts should proceed without further intervention.
This also extends to evaluation of the training and the resulting performance, where the necessary plots and statistics should be produced automatically.

Another advantage of this highly automatic training process is that it becomes possible for users running physics analyses to retrain the tag-side reconstruction for their own analysis.
Normally, if the software is trained on a generic $B$-decay Monte Carlo and reused for a certain signal-side selection, changes in the prior distribution may decrease classifier performance and introduce biases;
i.\,e. the trained methods assume combinatorial background from generic decays on the signal side, but receive selected data with much higher purity.
These differences should be especially notable for signal channels with only few tracks, e.g. $B \to \tau \nu$.
Retraining for a specific signal selection helps avoid these problems.

\section{Implementation in the Belle II software}
The Belle II analysis software framework (\texttt{basf2}) consists of independent modules mostly written in C++, which can be combined by the user using Python-based steering files~\cite{moll_basf2}.
For physics analyses, all common data-intensive tasks are performed by such modules and are provided via a common high-level syntax.
The following example demonstrates the usage of these tools in reconstructing $\pi^0 \to \gamma \gamma$ decays.
\begin{lstlisting}[language=Python]
selectParticle('gamma')

reconstructDecay('pi0 -> gamma gamma', '0.11 < M < 0.15')
matchMCTruth('pi0')
\end{lstlisting}
This selects final-state particles with a photon hypothesis, saves all possible combinations of two photons satisfying the given cut in a list of particle objects, and relates those objects to their Monte Carlo counterparts.

\begin{figure}
\begin{center}
\includegraphics[width=0.85\textwidth]{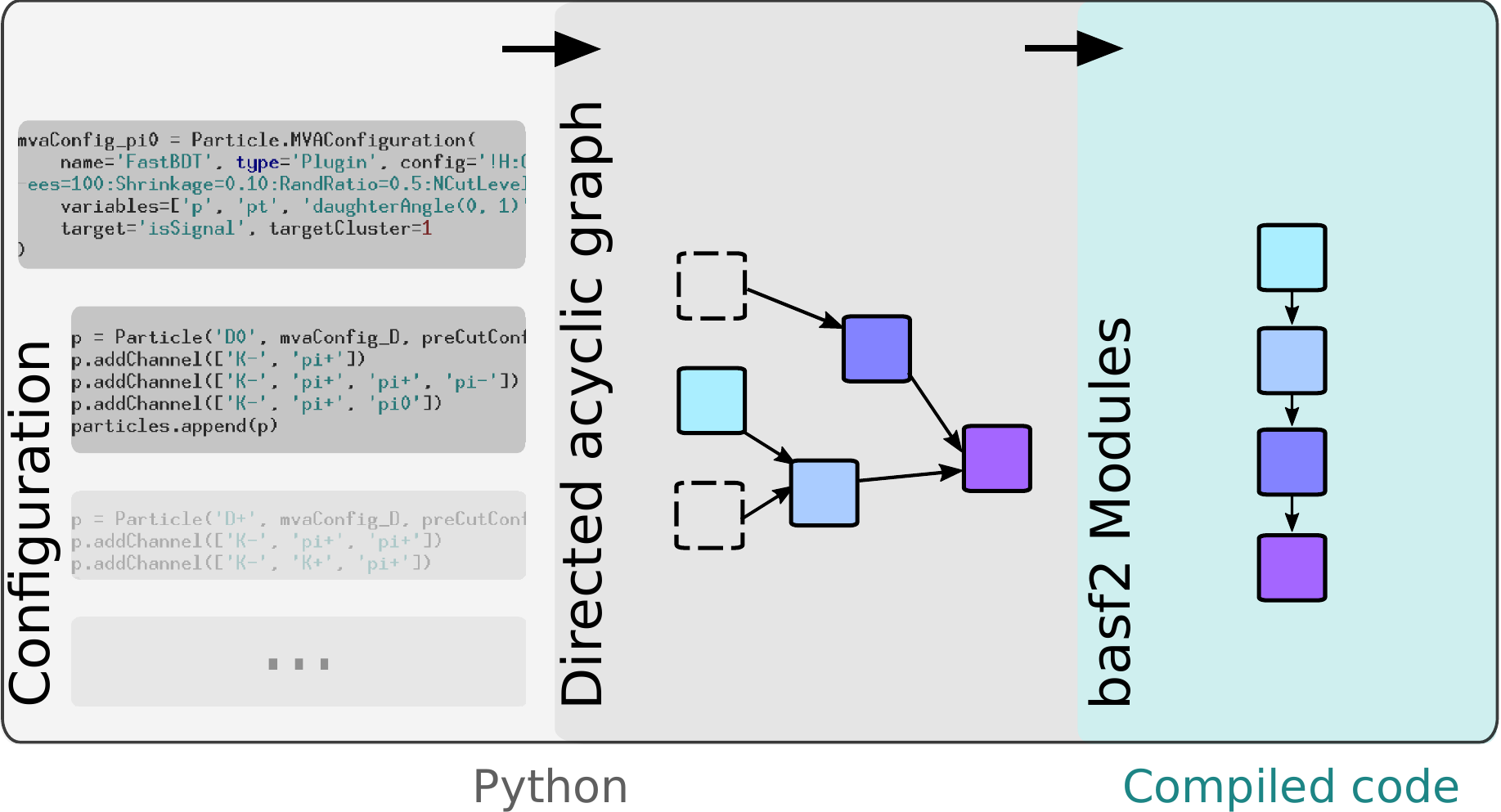}

\end{center}
\caption{\label{feiarchitecture}Architecture of the reconstruction framework's implementation, illustrating the configuration, dependency resolution between resulting reconstruction tasks, and the final data-intensive processing handled by basf2 modules.}
\end{figure}

To fully use these existing analysis tools, the hierarchical $B$ meson reconstruction framework itself is implemented in Python.
The main interface for the user is the configuration of particles and channels to reconstruct, as well as of the multivariate classifiers.
This high-level description is converted into separate reconstruction tasks, such as combining particles according to a specific decay channel, vertex fitting, or the training/application of a multivariate classifier.
Each of these tasks is represented by a functor object whose inputs define dependencies that need to be fulfilled by the output of other functors, or (user-supplied) configuration items.
Inputs and outputs thus produce a directed acyclic graph, which naturally defines an ordering in which functors should be executed to fulfill their dependencies.
Most functors do not directly process data, but only add modules to the execution path using the aforementioned reconstruction tools.
After the dependency resolution, the modules do data-intensive processing in the correct order.
This overall scheme is illustrated in Figure~\ref{feiarchitecture}.

For some tasks, outputs cannot be provided immediately.
For example, the multivariate classifiers provide a probability for a given candidate to be correctly reconstructed, but need to be trained first.
This is accomplished by checking for the existence of the training file and to only provide the output if it is detected.
If the file is not found, the training is done instead and execution of dependent tasks is deferred.
All file names and lists of particles include SHA1 hashes of the inputs used, so they will change once their configuration changes.
This in turn affects all subsequent functors, so the changes are propagated through the entire dependency graph.

\section{Technical details}
\subsection{Automatic cut determination}
\begin{figure}
\begin{center}
\begin{subfigure}{0.4\textwidth}
\includegraphics[width=\textwidth]{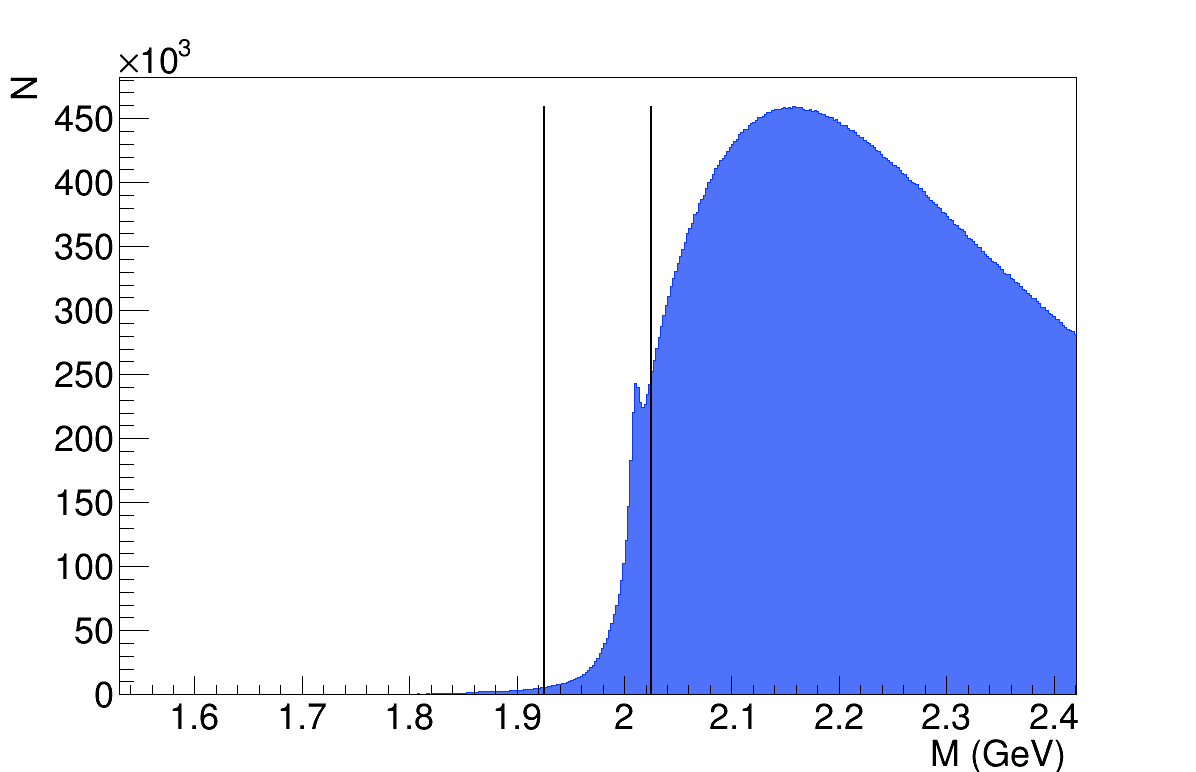}
\caption{All possible combinations}
\end{subfigure}
\begin{subfigure}{0.4\textwidth}
\includegraphics[width=\textwidth]{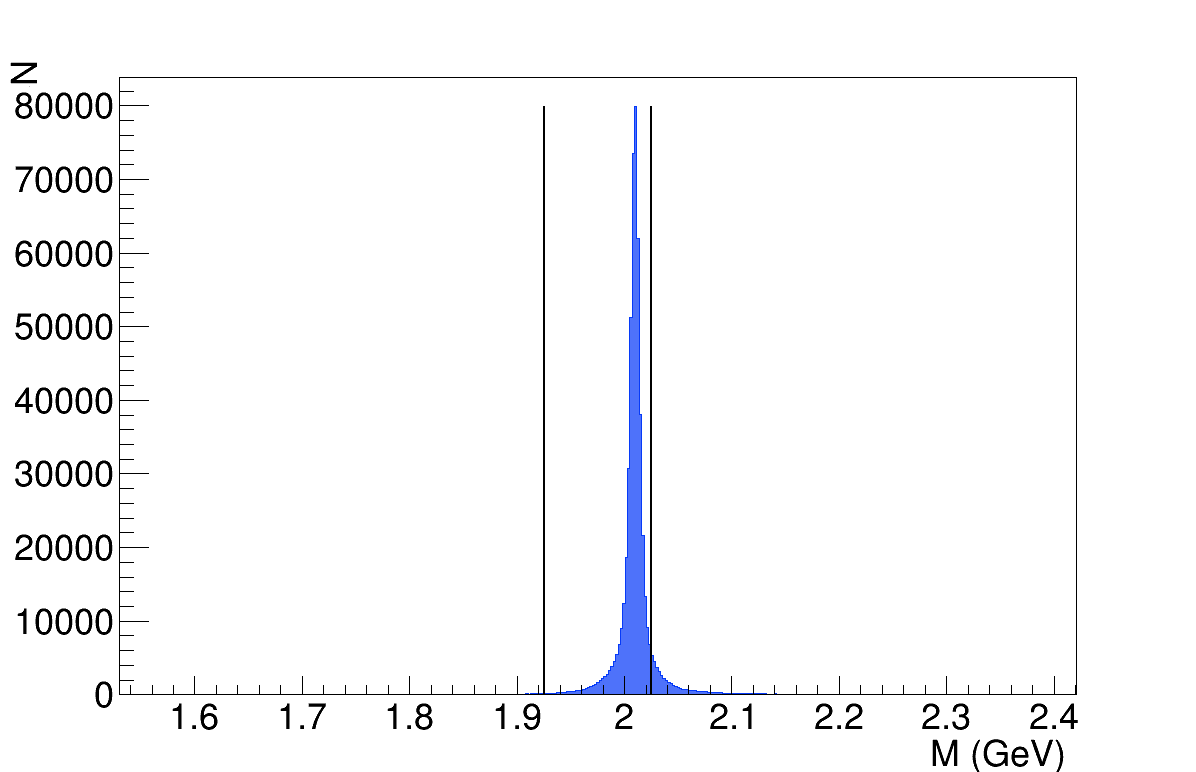}
\caption{Only signal combinations}
\end{subfigure}

\begin{subfigure}{0.4\textwidth}
\includegraphics[width=\textwidth]{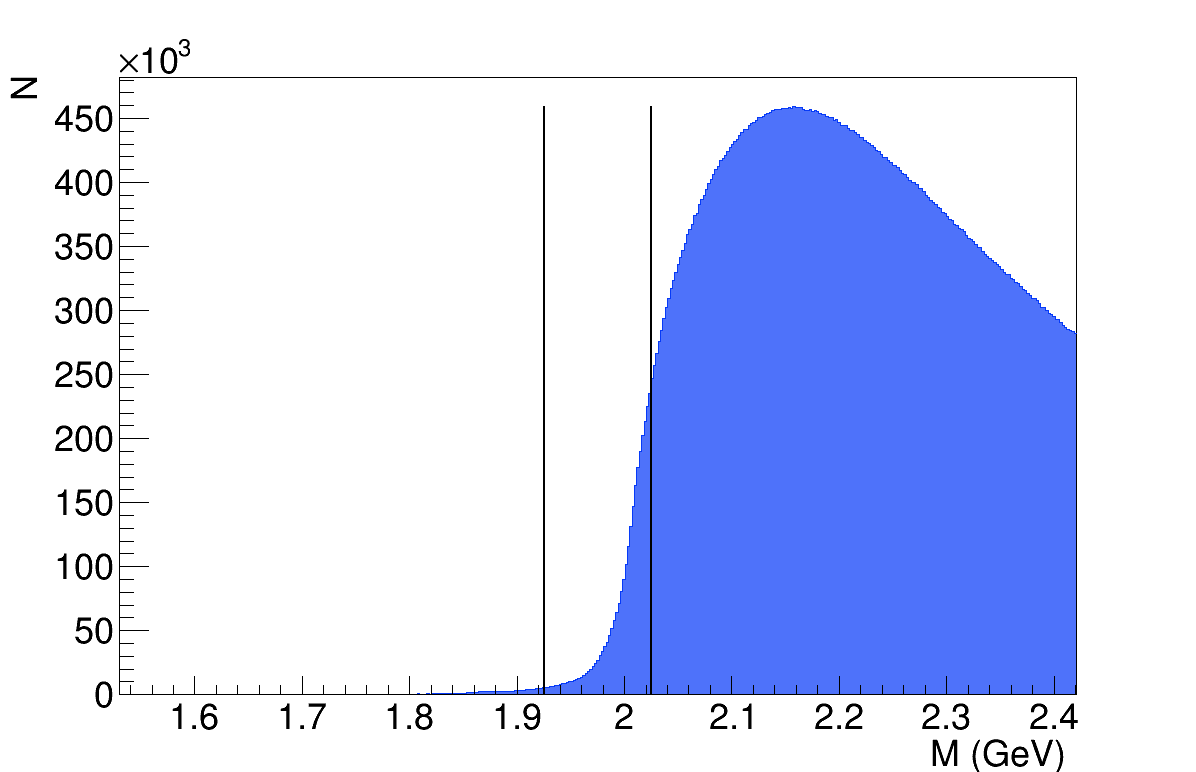}
\caption{Only background combinations}
\end{subfigure}
\begin{subfigure}{0.4\textwidth}
\includegraphics[width=\textwidth]{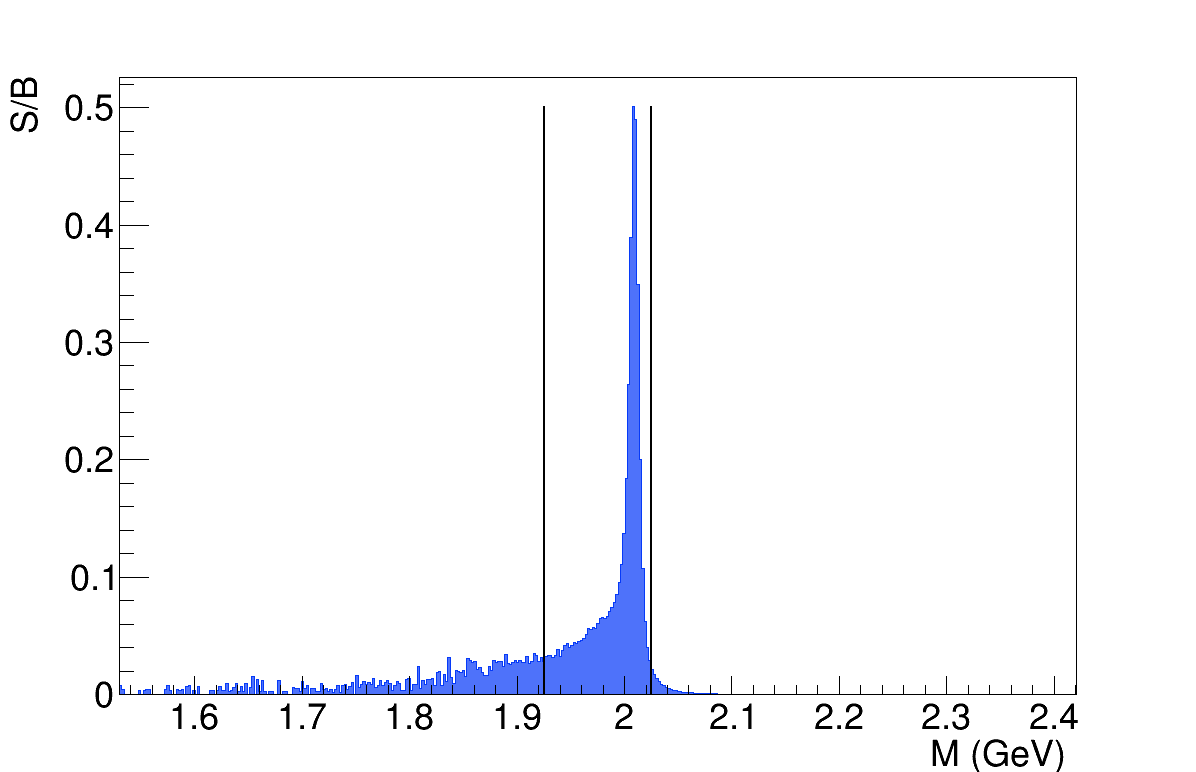}
\caption{Ratio $S/B$ of signal and background}
\end{subfigure}

\end{center}
\caption{\label{precuts}Plots illustrating the combinatorial background in the $D^{*+} \to D^0 \pi^+$ channel and its distribution in the $D^{*+}$ invariant mass $M$.
The lower signal-to-background ratio $S/B$ for higher masses automatically results in asymmetric pre-cuts (vertical lines).}
\end{figure}

To reduce the combinatorial background in the reconstruction, cuts need to be applied in between stages, before saving candidates.
Depending on the decay channel, we might cut on the invariant mass of a candidate, on the product of classifier output probabilities of the candidate's daughters, or any other variable.
Since the optimal cut often depends on the particular decay channel, requiring users to manually configure these cuts would discourage retraining or the addition of channels.
We address this by creating histograms of the signal and background distributions of the candidates in the cut variable and use them to automatically determine cuts that satisfy common requirements on the purity and efficiency.
Figure~\ref{precuts} shows pre-cuts determined for the $D^{*+} \to D^0 \pi^+$ channel as an example.
Channels in which the selection does not accept enough signal events to warrant the training of a multivariate classifier are excluded from further processing.

As a result, including potentially less clean channels poses no disadvantage to the process,
since they will either be assigned a cut that offers a useful compromise between the number of signal events and necessary CPU time, or the channels will be discarded entirely.

\subsection{TMVA interface}
Multivariate classification to discard incorrectly reconstructed particles is performed using TMVA, a multivariate analysis package for ROOT containing a number of different classification methods, pre-processing and evaluation tools~\cite{tmva}.
To make full use of the package inside basf2 two interface modules, \texttt{TMVATeacher} and \texttt{TMVAExpert}, are provided, which are meant to train or run classifiers on a list of particles.
Users can specify input variables, the classification target, and a list of methods that should be trained using module parameters, where variables are functions that return floating-point values for a candidate particle, e.g. the invariant mass of a particle or the $\chi^2$ of a vertex fit.
Some variables also take additional arguments to, e.g. return the angle between the candidate's momentum and that of its $i^\text{\tiny{th}}$ daughter, or to compute the product of a given variable for all daughter particles.
All variables are written in C++ and made available to all analysis modules through a string identifier.

\subsection{Automatic reporting}
To control the quality of the final $B_\text{tag}$ reconstruction, plots are generated automatically for each decay channel and its associated network.
This includes the pre-cut histograms shown in Figure~\ref{precuts}, plots provided by TMVA for the classifiers (receiver operating characteristic curves, overtraining checks, etc.), the ranking of input variables, and CPU usage statistics differentiated by channel and reconstruction task.
A table of the efficiencies and purities for each particle is also produced and allows inspection of the performance of the applied cuts.

\section{Summary}
Belle~II, as a $B$ factory, allows combining $B$ mesons selected as signal with a more generically reconstructed $B_\text{tag}$ to the original $\Upsilon(4S)$ resonance to improve many analyses.
As the usefulness of this technique depends strongly on the total efficiency of the $B_\text{tag}$ reconstruction, our framework makes it easy to add additional channels or optimize parts of the reconstruction.
This is achieved by automating many tasks, including the determination of intermediate cuts, the decision of which classifiers need to be retrained, and the generation of evaluation plots and tables.
The reconstruction framework itself is written in Python, the language used for basf2 steering files, and is built on top of high-level tools that can also be used by physics analyses.

The framework can be configured by defining particles and the channels in which to reconstruct them, plus related attributes like the configuration of multivariate methods or cuts.
It is generic enough to be reused for similar tasks like $B_\text{tag}$ reconstruction with semileptonic channels, or reconstructing $B_s$ mesons on $\Upsilon(5S)$ data samples.

\section*{References}
\bibliographystyle{iopart-num}
\bibliography{FEI_proceedings_acat2014}

\providecommand{\newblock}{}
\begin{thebibliography}{1}
\expandafter\ifx\csname url\endcsname\relax
  \def\url#1{{\tt #1}}\fi
\expandafter\ifx\csname urlprefix\endcsname\relax\def\urlprefix{URL }\fi
\providecommand{\eprint}[2][]{\url{#2}}

\bibitem{tdr}
{Abe} T {\em et~al.\/} (Belle II Collaboration) 2010 {Belle II Technical Design
  Report} Tech. rep. (\textit{Preprint} \eprint{1011.0352})

\bibitem{fullreco}
Feindt M, Keller F, Kreps M, Kuhr T, Neubauer S {\em et~al.\/} 2011 {\em
  Nucl.Instrum.Meth.\/} {\bf A654} 432--440 (\textit{Preprint}
  \eprint{1102.3876})

\bibitem{moll_basf2}
Moll A 2011 {\em Journal of Physics: Conference Series\/} {\bf 331} 032024

\bibitem{tmva}
Hoecker A, Speckmayer P, Stelzer J, Therhaag J, von Toerne E and Voss H 2007
  {\em PoS\/} {\bf ACAT} 040 (\textit{Preprint} \eprint{physics/0703039})

\end{thebibliography}

\end{document}